# Measuring Total Transverse Reference-free Displacements of Railroad Bridges using 2 Degrees of Freedom (2DOF): Experimental Validation

Lingkun Chen[1], Can Zhu[2], Zeyu Wu[3], Xinxing Yuan[4], and Fernando Moreu[5]

[1] Associate Professor, College of Civil Science and Engineering, Yangzhou University, 198 Huayang West Road, Yangzhou, Jiangsu, China,225127. Email: lkchen@yzu.edu.cn

[2] Graduate Student, College of Civil Science and Engineering, Yangzhou University, 198 Huayang West Road, Yangzhou, Jiangsu, China, 225127. Email: canzhu@unm.edu

[3] PhD, PE Associate Professor, School of Civil Engineering and Transportation, North China University of Water Resources and Electric Power, 36 Beihuan Road, Zhengzhou, Henan, China. Email: wuzeyu@ncwu.edu.cn

[4] PhD student, Department of Civil, Construction and Environmental Engineering, University of New Mexico, MSC01 1070, 1 University of New Mexico, Albuquerque, NM 87131. Email: xyuan@unm.edu

[5] Assistant Professor, Department of Civil, Construction and Environmental Engineering, Courtesy Appointment, Department of Electrical and Computer Engineering, Courtesy Appointment, Department of Mechanical Engineering, University of New Mexico, MSC01 1070, 1 University of New Mexico, Albuquerque, NM 87131. Email: fmoreu@unm.edu

**Abstract**

Railroad bridge engineers are interested in the displacement of railroad bridges when the train is crossing the bridge for engineering decision making of their assets. Measuring displacements under train crossing events is difficult. If simplified reference-free methods would be accurate and validated, owners would conduct objective performance assessment of their bridge inventories under trains. Researchers have developed new sensing technologies (reference-free) to overcome the limitations of reference point-based displacement sensors. Reference-free methods use accelerometers to estimate displacements, by decomposing the total displacement in two parts: a high-frequency dynamic displacement component, and a low-frequency pseudo-static displacement component. In the past, researchers have used the Euler-Bernoulli beam theory formula to estimate the pseudo-static displacement assuming railroad bridge piles and columns can be simplified as cantilever beams. However, according to railroad bridge managers, railroad bridges have a different degree of fixity for each pile of each bent. Displacements can be estimated assuming a similar degree of fixity for deep foundations, but inherent errors will affect the accuracy of displacement estimation. This paper solves this problem expanding the 1 Degree of Freedom (1DOF) solution to a new 2 Degrees of Freedom (2DOF), to collect displacements under trains and enable cost-effective condition-based information related to bridge safety. Researchers developed a simplified beam to demonstrate the total displacement estimation using 2DOF and further conducted experimental results in the laboratory. The estimated displacement of the 2DOF model is more accurate than that of the 1DOF model for ten train crossing events. With only one sensor added to the ground of the pile, this method provides owners with approximately 40% more accurate displacements.

## 1. Introduction

Structural assessment using sensors is increasing as a reliable method to inform railroad bridge management and to prioritize replacements. In the past, researchers determined the damage of existing in-service structures using sensors. In their work, damage is defined by changes in the geometric properties and materials of the system. Structural Health Monitoring (SHM) analyzes and processes data from the structures to provide early warnings and/or to inform about damage or damage growth. This includes changes in the system connectivity and boundary conditions that have an adverse effect on the health of the structure (Farrar, 1998, 2001, 2006). Researchers are interested to use sensors to understand the response of structures (Brownjohn, 2007). Sensors can play an important role to locate damage in real-time structure monitoring systems (Hannan, 2018) by sending measurements to the servers on the Internet via standard protocols, which can be analyzed and inform owners of changes in the structure remotely (Chang and Lin, 2019). Pandey et al. (1995) introduced the application of multilayer perception in steel bridge structural damage detection. Zhu et al. (2007) studied the effects of axle system parameters and measurement of noise on the damage detection results. S. Park et al. (2006) presented an experimental study based on Piezoelectric lead-Zirconate-Titanate (PZT) active damage detection technology for non-destructive evaluations (NDE) of steel bridge components. Salawu (1997) discussed the use of natural frequencies as diagnostic parameters in structural evaluation procedures for vibration monitoring. Wang et al. (2019) estimated the vehicle dynamic responses crossing a bridge under wind with the hybrid combination of SHM data and a dynamic simulation of single vehicle model. Galvín et al. (2021) described the main structural characteristics that affect the dynamic characteristics of the bridge, determining the importance of knowing the soil properties of the structure at each location to analyze the dynamic response of the bridge under different operating conditions. These approaches are directed to find damage identifying changes in the structural properties using models. However, stakeholders are averse to collect data in the field that needs to be related to models, since every bridge is different and not all their structural properties are available. Railway managers want to quantify the structural performance of bridges in the field in real-time without having to know the properties of their bridges (model-free assessment) (Moreu et al. 2016).

Managers are interested to use field data related to hazardous performance to inform bridge management systems. Pregnolato (2019) pointed out the importance of developing bridge management systems informed by data, and introduced a risk-based bridge management method with a preliminary pattern classification method in flood prone areas.

Owners of railroad bridges objectively prioritize which components and bridges to repair cost-effectively within their network using displacement amplitude and displacement changes across time (Moreu et al. 2017). Structural displacement can indicate structural safety of railroad operations and can inform decisions on management (Moreu et al. 2014). Linear Variable Differential Transformer (LVDT) is the basic sensor of displacement measurement but cannot be effectively applied in the field due to lack of references from where to measure from (Park et al. 2005; Moreu et al. 2016). Researchers have developed indirect displacement estimation methods to overcome the difficulties associated with direct measurement methods in the field. Indirect methods use accelerations, strains, and/or structural properties (Gindy et al. 2008; Park et al. 2013; Helmi et al. 2015). Researchers measured total bridge displacement using sensor fusion and structural properties (Park et al. 2014). Shin et al. (2012) created an algorithm for estimating the vibration displacement of a bridge using the vibration strain measured using Fiber Bragg Grating (FBG) sensors and a simply supported beam model. These methods rely on structural strength knowledge and multiple sensors using those properties to estimate displacements and models of the structure. Owners are interested to add sensors in their bridges to inform objective management but are not interested in building one model for each bridge being monitored as that takes time, resources, and costs. To solve the aforementioned problems, Ozdagli et al. (2018) used low-cost sensors to measure reference-free, model-free displacements of simplified models of piles and columns without structural properties. The errors from their experimental validation were in general under 10% for both peak to peak displacement and RMS measurements assuming a cantilever beam with 1 Degree of Freedom (1DOF). According to the railroad, it would be advantageous to increase their level of accuracy by reducing errors in estimation, while maintaining the reference-free, model-free approach.

    This paper designs and investigates the improvement of accuracy of a reference-free displacement estimation method without knowledge of structural properties by using 2 Degrees of Freedom (2DOF) instead of 1DOF. The proposed method provides accurate displacement estimation data, especially for timber railway bridges that experience pseudo-static deflection due to asymmetric boundary conditions (i.e. different foundation properties.) The railroad is interested to prioritize the replacement of existing timber bridges to improve safety in their network and to prioritize management decisions using accurate displacement. The research team validated the method conducting ten shake table experiments simulating actual traffic conditions corresponding to North American timber railway bridges. Then, researchers modified the Euler-Bernoulli beam theory formula from 1DOF to 2DOF to estimate the pseudo-static displacement. The results from the 2DOF method were compared to those obtained using the 1DOF method and validated with a conventional LVDT displacement sensor in the laboratory. The conclusions of this study show that the proposed 2DOF method can estimate lateral displacement of railway

bridges 40% more accurately than the 1DOF method. The proposed 2D method is simple and does not require structural properties of the timber bridge nor foundation information, plus the new sensor in at the ground level (accessible to inspectors.) This method improves the ability of railway bridge managers to make informed decisions while maintaining the low-cost, low-effort merit that is required for practical informed management of civil infrastructure.

## 2. Principles of Displacement Estimation

If displacements could be easily collected, railroad owners could make decisions on their management programs prioritizing actions on bridges with worst displacements under revenue service traffic (Moreu et al. 2014.) Railway bridges that are susceptible to asymmetric loads causing lateral displacement composed of both dynamic and pseudo-static components. Offset track can cause during the train-crossing event an eccentric vertical load (Dias, 2007; Lee et al. 2005). Ozdagli et al. (2017) showed that this eccentric vertical load can cause a pseudo-static lateral deflection of the bridge, creating a low-frequency displacement as a result of this asymmetric boundary condition. Ozdagli et al. (2017) were able to estimate displacements using 1DOF assumptions, in order to obtain a reference-free, model-free displacement. Aguero et al. (2019) estimated transverse displacements with the same method experimentally using low-cost wireless sensors. However, estimating total displacement more accurately using MDOF can benefit railroads and their decisions by reducing the error without needing to know changes in foundation properties. The following sections provide a background on the estimation of dynamic displacement and the 1DOF pseudo-static displacement to introduce the 2DOF solution.

### 2.1. Dynamic displacement estimation: background

This section summarizes how to estimate the dynamic displacement. Park et al. (2005), Moreu et al. (2016) and Hester et al. (2017) reported that the dynamic response is in general above 0.5 Hz.

The FIR filter is used to calculate the measured acceleration and derive the zero average dynamic displacement (Lee et al. 2010), since the boundary conditions are not accurate and the double integral of the acceleration increases the displacement drifts. Lee et al. method estimates the dynamic displacement with:

$$\Delta_d = (L^T L + \lambda^2 I)^{-1} L^T L_a \bar{a}(\Delta t)^2 = C\bar{a}(\Delta t)^2 \quad (1)$$

where $\Delta_d$ = estimated dynamic displacement; $L$ = diagonal weighting matrix; $\lambda$ = optimal regularization factor; $I$ = unit matrix of (2k+3) order; $L_a$ = diagonal weighting matrix of (2k+1) order; $\bar{a}$ = measured acceleration; $\Delta t$ = time increment; and $C$ = displacement estimation coefficient matrix.

Additionally Lee et al. (2010) derived the optimal regularization factor, defined as

$$\lambda = 46.81 N^{-1.95} \quad (2)$$

where $N$ = the number of points corresponding to the time windows. Researchers have used Lee et al.'s method demonstrating its ability of FIR filter to estimate dynamic displacement: Moreu et al. (2016) and Park et al. (2014; 2016).

### 2.2.1DOF total displacement estimation: background

The total displacement requires both dynamic and pseudo-static estimation. The pseudo-static displacement estimation has been calculated in many applications, such as mobile phone and virtual reality headsets orientation (Fisher, 2010). Ozdagli et al. (2017, 2018) and Aguero et al. (2019) estimated the pseudo-static displacement from the inclination angle using accelerometers. The pseudo-static response usually ranges from 0 to 0.5 Hz for North American railroad bridges Ozdagli et al. (2017, 2018.) The sensor measures the acceleration on the $x$ axis. When the sensor is inclined, the projection of the gravitational acceleration, g, produces an output acceleration $A_x$ on the $x$ axis of the sensor (Figure 1) (Ozdagli et al. 2018.)

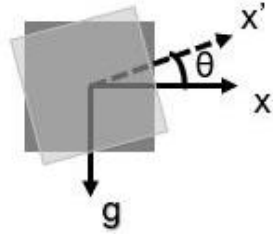

Figure 1. Single axis accelerometer sensing rotation diagram.

The following equations define how to obtain the pseudo-static displacement using accelerations.

$$A_x = g * \sin(\theta) \tag{3}$$

$$\theta = \sin^{-1}\left(\frac{A_x}{g}\right) \tag{4}$$

$$R = g * (\sin(N + P) - \sin(N)) \tag{5}$$

where $R =$ the minimum required resolution; $N =$ the angle range to be measured; and $P =$ the minimum measuring angle (Fisher, 2010).

Two single axis accelerometers at right angles to each other are combined to get accurate data (Figure 2). The cosine of the angle θ between the gravity vector and the $y$ axis is calculated converting the measured $y$ axis acceleration $A_y$ into an inclination angle (Ozdagli et al. 2018).

$$A_y = g * \cos(\theta) \tag{6}$$

$$\theta = \cos^{-1}\left(\frac{A_y}{g}\right) \tag{7}$$

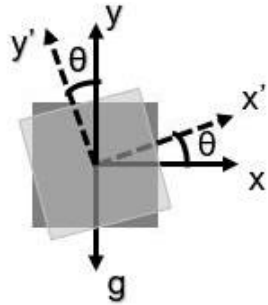

Figure 2. Two axis accelerometer sensing rotation diagram.

By combining Eq. (3) and Eq. (6), the ratio between $A_x$ and $A_y$ can be calculated, which can be used to determine the tangent of the inclination:

$$\frac{A_x}{A_y} = \frac{g*\sin(\theta)}{g*\cos(\theta)} = tan(\theta) \tag{8}$$

$$\theta = tan^{-1}(\frac{A_x}{A_y}) \tag{9}$$

Researchers used a simple moving average (SMA) filter to obtain the inclination angle of the pseudo-static component:

$$\theta_1[i] = \frac{1}{n}\sum_{j=0}^{n-1} \theta_t[i+j] \tag{10}$$

where $\theta_t$ = the total inclination angle; $i$ = $i^{th}$ time step; $n$ = the average number of points.

The following section develops the new 2DOF method and investigates its accuracy vs. that of the 1DOF method. Researchers used ten experiments to validate experimentally the new approach for simplified, objective management of railroad bridges.

### 3. Derivation of the 2DOF Displacement Estimation from 1DOF

Figure 3 describes the structural assumptions and context for the new algorithm. Figure 3(a) shows a typical American timber railway bridge bent and its cross-sectional view. Figure 3(b) shows the transverse deformation under train crossings. The lateral section of the bridge is simplified to a cantilever column, which is the main part of the 1DOF, as illustrated in Figure 3(c). In the 1DOF method, $\theta_2$ is assumed to be zero to simulate a cantilever column with one end fixed and no angle at the support. Assuming load P producing pseudo-static displacement $\Delta_P$, the resulting pseudo-static rotation at the top of the column is equal to $\theta_1$. According to the Euler-Bernoulli beam theory, $\Delta_P$ and $\theta_1$ are interrelated independently of $P$, $E$, and $I$, as follows:

$$\Delta_p = \frac{PL^3}{3EI} \tag{11}$$

$$\theta_1 = -\frac{PL^2}{2EI} \tag{12}$$

$$\Delta_p = -\frac{2}{3}\theta_1 L \tag{13}$$

where $\Delta_p$ = pseudo-static displacement; $P$ = lateral load; $L$ = the length of cantilever column; $I$ = the moment of inertia of column section; $E$ = Young's modulus; $\theta_1$ = the resulting pseudo-static rotation at the top of the column.

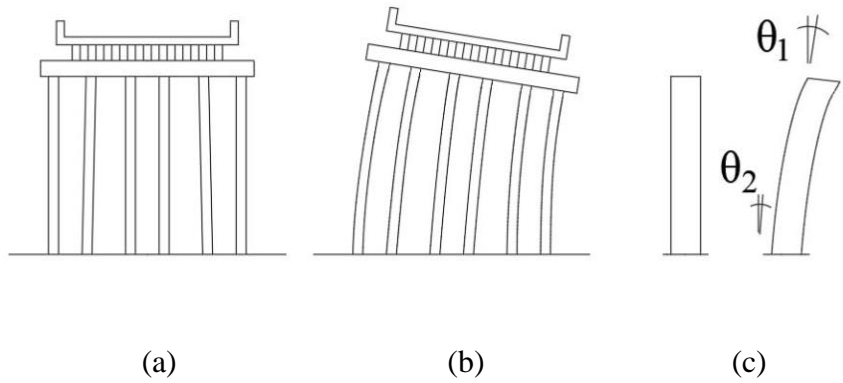

(a)         (b)         (c)

Figure 3. Cantilever simplification for a bridge.

However, in actual bridges, the bottom of the bridge is not fixed and has an angle $\theta_2$ non-zero. This new method proposes a 2DOF using two sensors located between the support and the top of the pile as shown in Figure 4. If load P produces a pseudo-static displacement, $\Delta_p$, the pseudo-static rotation generated at the free end is equal to $\theta_1$ and the pseudo-static rotation generated at the support is $\theta_2$.

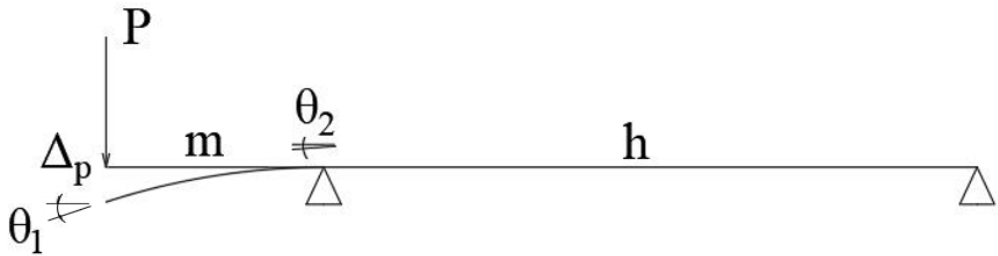

Figure 4. Simplification of the new pile.

With the new 2DOF, both Eq. (11) to Eq. (13) need to be redefined. According to the Euler-Bernoulli beam theory, $\Delta_p$, $\theta_1$ and $\theta_2$ can be estimated as follows:

$$\lambda = \frac{m}{h} \tag{14}$$

$$\Delta_p = \frac{Pm^2h}{3EI}(1 + \lambda) \tag{15}$$

$$\theta_1 = -\frac{Pmh}{6EI}(2 + 3\lambda) \tag{16}$$

$$\theta_2 = -\frac{Pmh}{3EI} \tag{17}$$

where $m$ = length of the pile above ground; $h$ = length of the pile below ground; $\lambda$ = ratio of the length of the pile above ground and the pile segment below ground; $P$ = load; $I$ = the moment of inertia of the beam; $E$ = Young's modulus; $\Delta_p$ = pseudo-static displacement; $\theta_1$ = the resulting pseudo-static rotation at the top of the pile; $\theta_2$ = the resulting pseudo-static rotation at the bottom of the pile.

By combining Eq. (16) and Eq. (17), the relationship of $\theta_1$, $\theta_2$ and $\lambda$ can be obtained.

$$\lambda = \frac{2}{3}\left(\frac{\theta_1}{\theta_2} - 1\right) \tag{18}$$

Further substituting $\lambda$ into the following equation to find the angle relationship to the overhang:

$$\frac{\Delta_p}{\theta_1} = \frac{\frac{Pm^2h}{3EI}(1+\lambda)}{-\frac{Pmh}{6EI}(2+3\lambda)} = -2m\frac{1+\lambda}{2+3\lambda} = -\frac{2}{3}m\left(1 + \frac{1}{2+3\lambda}\right) = -\frac{2}{3}m - \frac{m\theta_2}{3\theta_1} \tag{19}$$

Therefore, the relationship between pseudo-static displacement and angles depends only of the length of the pile above ground, and the two rotations collected with two sensors. More importantly, the total pseudo-static displacement is independent of the foundation properties, the pile length below ground, and the external load, as shown in Eq. (20):

$$\Delta_p = -\frac{2}{3}\theta_1 m - \frac{1}{3}\theta_2 m \tag{20}$$

## 4. 2DOF Displacement Estimation Validation

The total estimated displacement can be obtained by superimposing the dynamic and pseudo-static estimation displacements.

$$\Delta_t = \Delta_d + \Delta_p \tag{21}$$

where $\Delta_t$ = the total estimated displacement; $\Delta_d$ = the dynamic estimated displacement; $\Delta_p$ = the pseudo-static estimated displacement. Figure 5 summarizes the flowchart for the comparison of the two methods.

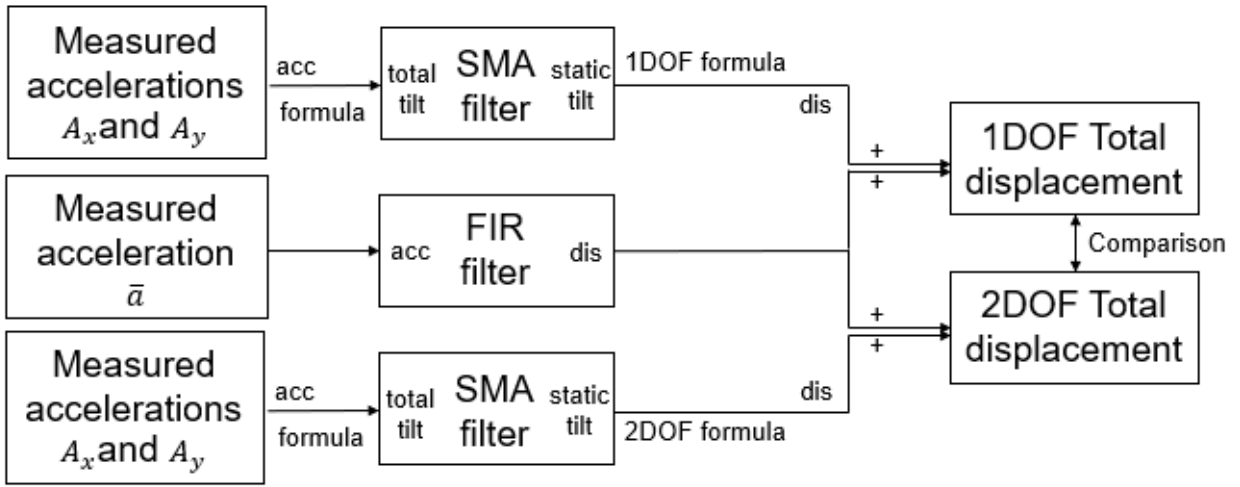

Figure 5. Flowchart comparison between model-free, reference-free 1DOF and 2DOF total displacement estimation.

The total displacement estimation with the new 2DOF method is divided into four parts: (1) dynamic estimated displacement are calculated from the acceleration $\bar{a}$ using the FIR filter; (2) the pseudo-static estimated displacements are obtained from Eq. (13) (1DOF formula) and Eq. (20) (2DOF formula) from accelerations $A_x$ and $A_y$ using Eq. (9) and the SMA filter; (3)

1DOF and 2DOF total displacements are obtained adding the dynamic and the pseudo-static displacements; and (4) 2DOF and 1DOF displacements are compared vs. LVDT displacements.

### 4.1. Experiment Instrumentation

Researchers conducted their experiments using a shake table to simulate ten train-crossing events. Researchers used accelerometer and LVDT for collecting acceleration data and displacement data, respectively. The capacitive accelerometers are all manufactured by PCB Piezotronics, Model 3711E1110G D. These models all have a strong, fully welded titanium alloy housing; internal voltage regulators; built-in microelectronics technology; a gas damper that extends the higher frequency range and reduces unnecessary high frequency vibrations; and a sealed multi-pin connector. These accelerometers provide reference-free data acquisition for the 2DOF method. The LVDT consists of a primary coil, two secondary coils, a core, a coil bobbin, and a casing. The LVDT provides: (1) frictionless measurement; (2) unrestricted resolution; and (3) input/output isolation. The LVDT provide ground truth data for validation. The proposed method will only use reference-free accelerometers in the field.

### 4.2. Experimental Setup

Figure 6 shows the experimental setup, replicating pile vibration under train events. The research team used a shake table to replicate the pile transverse response generated during a train-crossing event (Figure 6a). The pile specimen is 44.7 cm long, 14.8 cm wide and 0.8 cm thick. The modulus of elasticity and the rigidity are $0.92 \times 10^9$ Pa and 104.7548 N/m, respectively.

The specimen captures the 2DOF configuration of the railroad pile. The specimen is installed upside down, where its base moves with the shake table, as the top of a railroad pile under train crossing events. The base of the specimen is able to move right and left under vibrations as a hinge. The upper end of the specimen represents the bottom of the railroad pile, able to rotate with partial fixity, replicating field situations. To ensure the partial fixity only one U-shaped steel piece connects the specimen to the frame (Figure 6b).

Throughout the experiment, the researchers need to collect acceleration data from the top and bottom of the pile for 1DOF and 2DOF for estimation, and displacement data at the top of the pile for validation. Two capacitive accelerometers are attached to the bottom (Figure 6c) and the top (Figure 6d) of the specimen, two capacitive accelerometers on each level. The LVDT measure the displacement data for validation (Figure 6e). Researchers used this experimental setup to estimate total reference-free displacement using both 1DOF and 2DOF methods for validation and comparison.

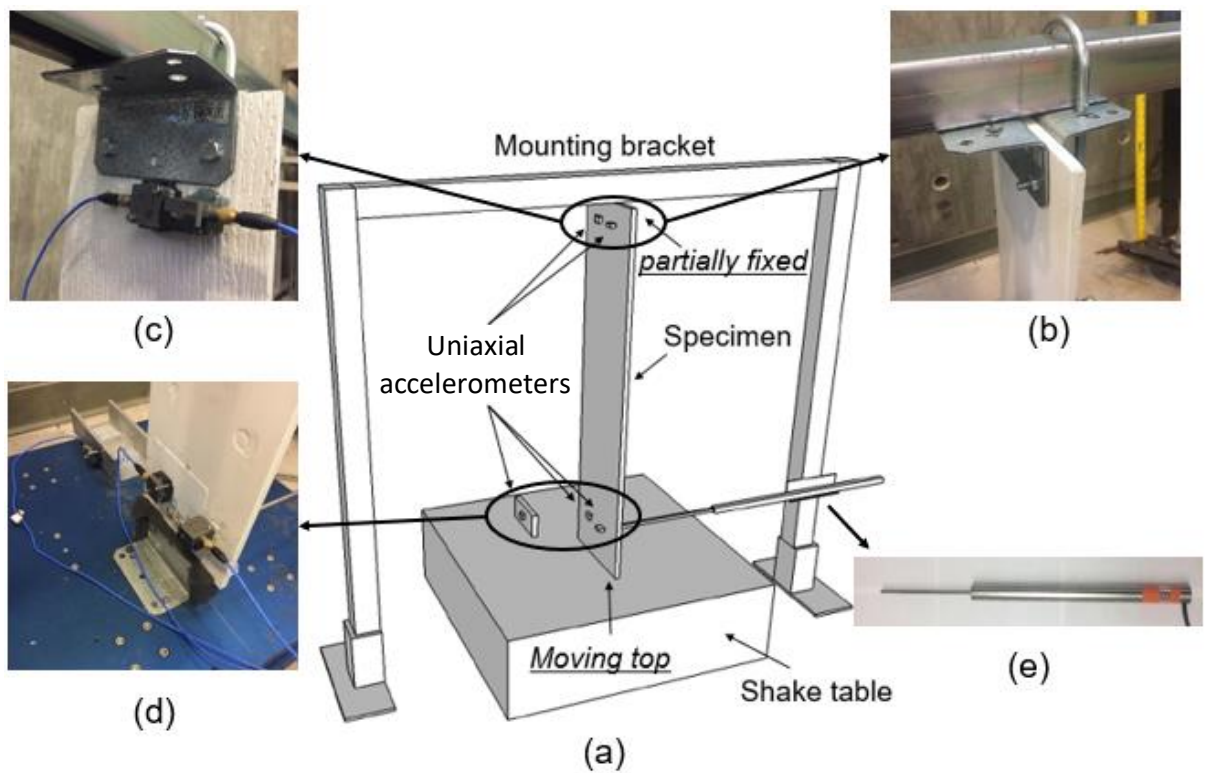

Figure 6: Experiment details: (a) Overall experimental setup; (b) Details of partially fixed support; (c) Details of bottom sensors ($\theta_2$); (d) Details of top sensors ($\theta_1$); and (e) LVDT.

### 4.3. Input Displacement Description

Researchers input the displacement time history data of the total displacement measured during ten train crossings events of at a railroad timber bridge, as measured by Moreu et al. (2016) (Table 1, Figure 7). Researchers used the variability of displacements, rotations, speeds, and directions to better test the accuracy of the 2DOF vs. the 1DOF, using the LVDT as reference. The ten train crossing events have different speeds and duration. The absolute maximum amplitudes of Train 1 and Train 10 are 6.273 mm and 13.925 mm, respectively.

Table 1. Ten sets of displacement time history data details.

| Train Number | Duration (seconds) | Maximum amplitude (mm) | Minimum amplitude (mm) | Train speed [km/h (mph)] |
|---|---|---|---|---|
| Train 1 | 76.00 | 1.563 | -6.273 | 8.7 (5.4) |
| Train 2 | 74.82 | 2.633 | -6.506 | 8.7 (5.4) |
| Train 3 | 34.56 | 1.301 | -8.324 | 16.2 (10.1) |
| Train 4 | 33.73 | 4.075 | -8.208 | 17.8 (11.0) |
| Train 5 | 25.21 | 4.970 | -7.134 | 23.3 (14.5) |
| Train 6 | 20.33 | 9.855 | -11.058 | 24.9 (15.5) |
| Train 7 | 28.89 | 4.873 | -8.154 | 33.9 (21.0) |
| Train 8 | 16.29 | 13.700 | -15.381 | 31.1 (19.3) |
| Train 9 | 13.36 | 5.656 | -12.441 | 41.5 (25.8) |
| Train 10 | 11.29 | 13.925 | -12.32 | 41.0 (25.5) |

### 4.4. Experimental Results

Researchers collected the acceleration data for dynamic displacements converting $A_x$ and $A_y$ into inclination angle. To compare the pseudo-static displacements between 1DOF and 2DOF methods, the LVDT data is corrected using an SMA filter. Researchers also obtained a dynamic reference displacement subtracting the pseudo-static reference displacement obtained from the SMA filter from the reference displacement.

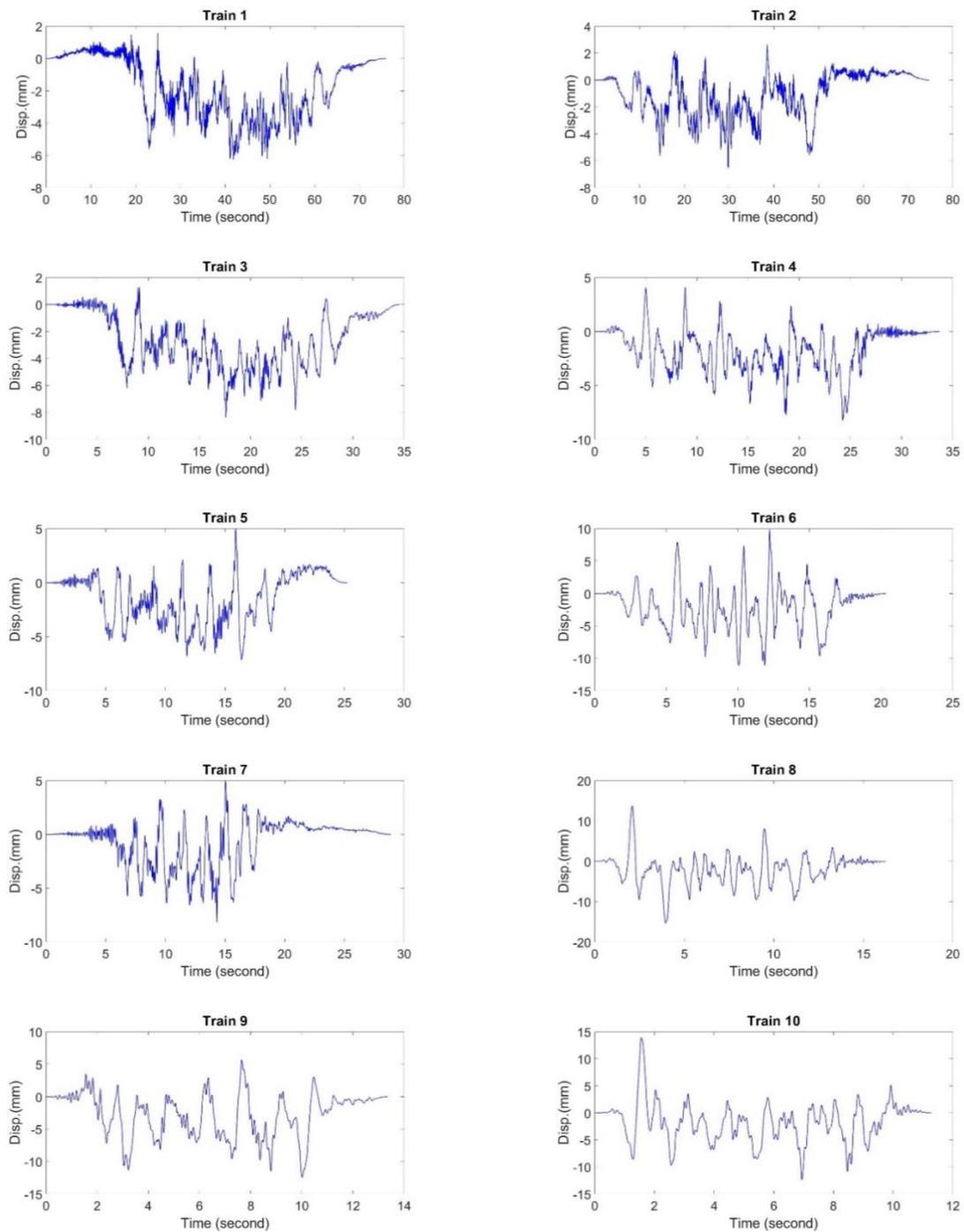

Figure 7. Summary of transverse displacements from ten train-crossing events from the field used for the laboratory experiments.

Figure 8 compares reference-free dynamic displacements from both 1DOF and 2DOF methods and the LVDT measured dynamic displacement. Both 1DOF and 2DOF methods obtain the same dynamic displacement. Dynamic displacements errors are the same for both 1DOF and 2DOF.

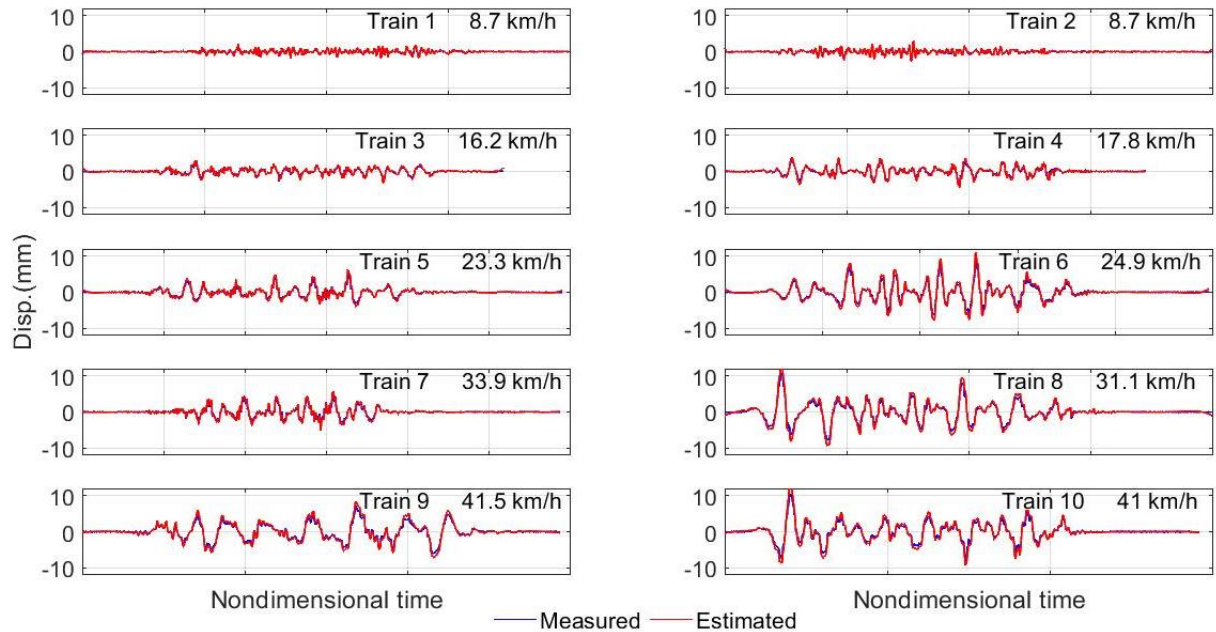

Figure 8. Comparison of dynamic displacement estimation and displacement measurement.

Figure 9 compares pseudo-static displacements from 1DOF and 2DOF methods and the LVDT measured pseudo-static displacements. In this case, both 1DOF and 2DOF pseudo-static displacements are different. In general, pseudo-static displacements from the 2DOF method are more accurate than those of the 1DOF when compared with the LVDT pseudo-static measurements. Figure 10 compares total displacements from 1DOF and 2DOF methods and the LVDT measured total displacements. In general, total displacements from the 2DOF method are more accurate than those of the 1DOF when compared with the LVDT pseudo-static measurements.

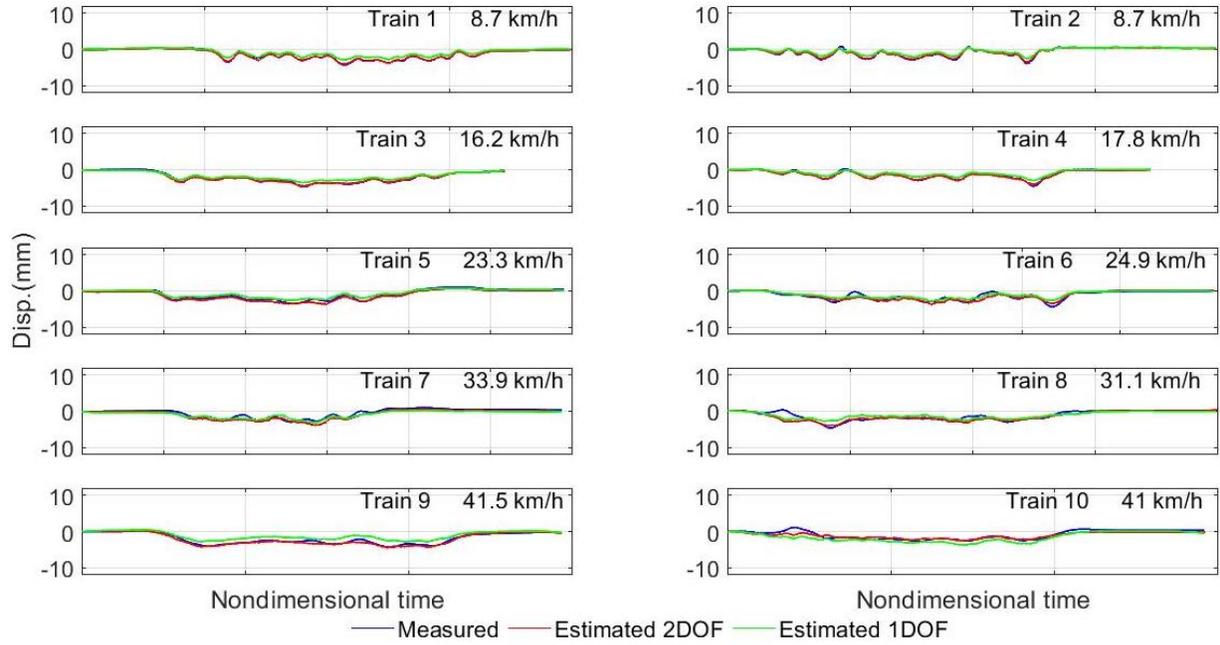

Figure 9. Pseudo-static displacement estimation with 1DOF and 2DOF methods.

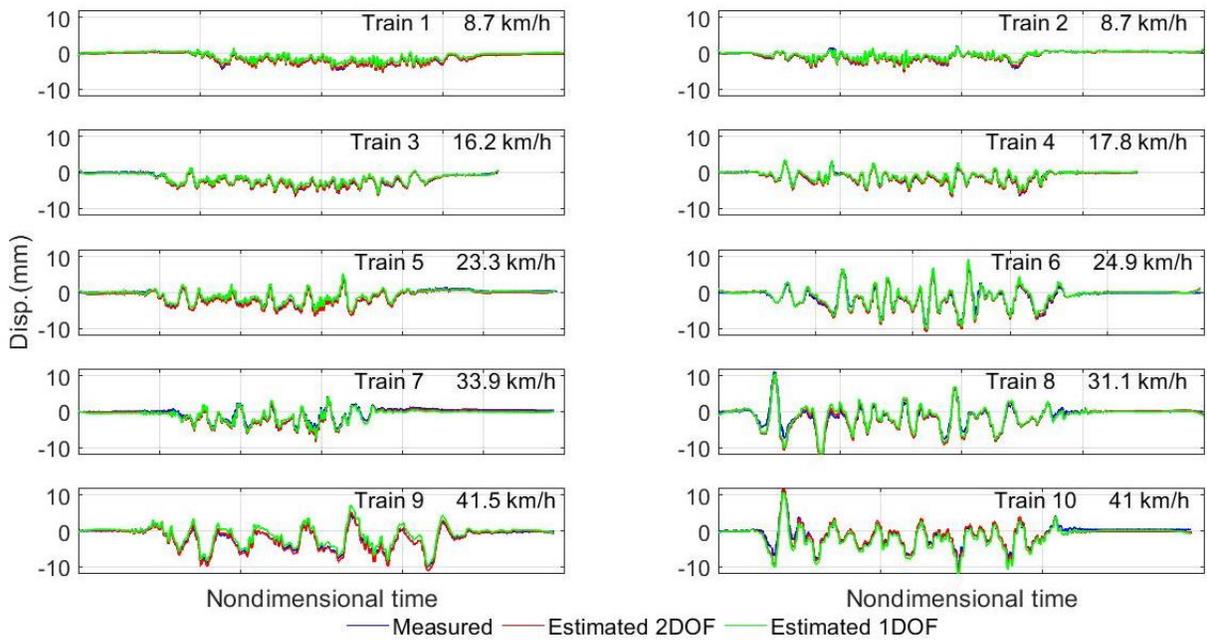

Figure 10. Total displacement estimation with 1DOF and 2DOF methods.

Researchers quantified the accuracy of both displacement estimations using two indexes. The first index $E_1$ determines the peak error between the estimated displacement and the actual displacement. The second index $E_2$ determines the root mean square (RMS) error between the estimated displacement and the actual displacement. Both indexes were normalized in order to enable relative comparison of errors. $E_1$ and $E_2$ are defined as follows:

$$E_1 = \frac{|\Delta_{est}|_{max} - |\Delta_{meas}|_{max}}{|\Delta_{meas}|_{max}} \quad (22)$$

$$E_2 = \frac{RMS(\Delta_{est} - \Delta_{meas})}{RMS(\Delta_{meas})} \quad (23)$$

where $\Delta_{est}$ = estimated displacement; and $\Delta_{meas}$ = actual displacement measured $RMS(i)$ = the root mean square of (i).

Table 2 summarizes the performance of both methods and both indexes. A smaller index indicates a better performance. In all ten trains and for both indexes, the 2DOF total displacement estimation is closer to the LVDT measurement. The results of the ten experiments show that the E$_1$_2DOF error is less than 10% for 9 of the 10 train-crossing events, with the maximum error being 11.8% under Train 6. The average value of E$_1$_2DOF is 5.03%, which is less than 9.60% of E$_1$_1DOF. Similarly, the E$_2$_2DOF errors for all ten experiments are below 8%, and the maximum E$_2$_2DOF error is 7.9% under Train 5. The average value of E$_2$_2DOF is 5.45%, which is less than 8.51% of E$_2$_1DOF. On average, the 2DOF method reduces the error of displacement estimation in peak to peak and RMS estimations by 48% and 36%. A general improvement of accuracy of 40% in displacement estimation is a significant improvement by only adding one new sensor in each pile at the ground level of the railroad bridge. The maximum error of E$_1$_2DOF corresponding to the peak error was 11.8% and the maximum error of E$_2$_2DOF corresponding to the RMS error was 7.9%, as opposed to 16.6% and 11.4% for the 1DOF indexes, 41% and 44% less accurate, respectively. Figure 11 summarizes the two errors for all ten trains.

Figure 11(a) and Figure 11(b) show that the error values obtained are smaller for the 2DOF for both $E_1$ and $E_2$, respectively for all experiments. The results obtained when considering the angle of the base of the cantilever column (2DOF) are better than those obtained by ignoring it (1DOF). In the future stages of this research, the authors will explore MDOF for more precision in their estimations. Additionally, the authors will also consider non-linearities attributed to damage in the pile, which were not included in this algorithm.

Table 2. Performance result.

| Train number | $E_1$_1DOF (%) | $E_1$_2DOF (%) | $E_2$_1DOF (%) | $E_2$_2DOF (%) |
|---|---|---|---|---|
| Train 1 | 16.6 | 2.4 | 11.4 | 3.6 |
| Train 2 | 10.3 | 4.5 | 9.9 | 4.0 |
| Train 3 | 10.2 | 2.6 | 8.0 | 3.8 |
| Train 4 | 7.0 | 5.1 | 7.6 | 4.3 |
| Train 5 | 10.4 | 8.2 | 8.2 | 7.9 |
| Train 6 | 13.2 | 11.8 | 6.7 | 6.6 |
| Train 7 | 9.4 | 8.2 | 8.0 | 7.8 |
| Train 8 | 4.0 | 3.7 | 6.8 | 6.4 |
| Train 9 | 4.3 | 3.7 | 10.1 | 4.4 |
| Train 10 | 10.6 | 0.1 | 8.4 | 5.7 |
| Average value | 9.60 | 5.03 | 8.51 | 5.45 |

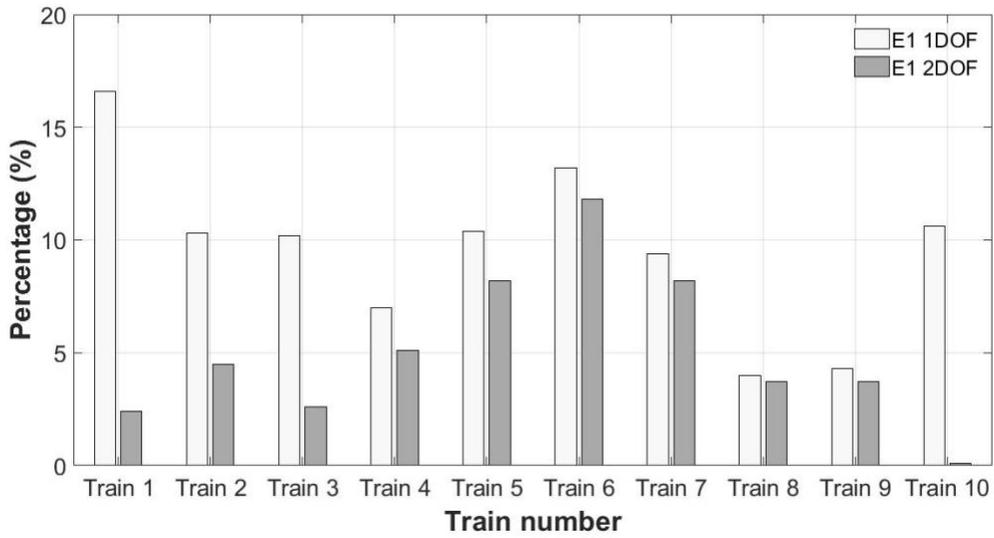

(a)

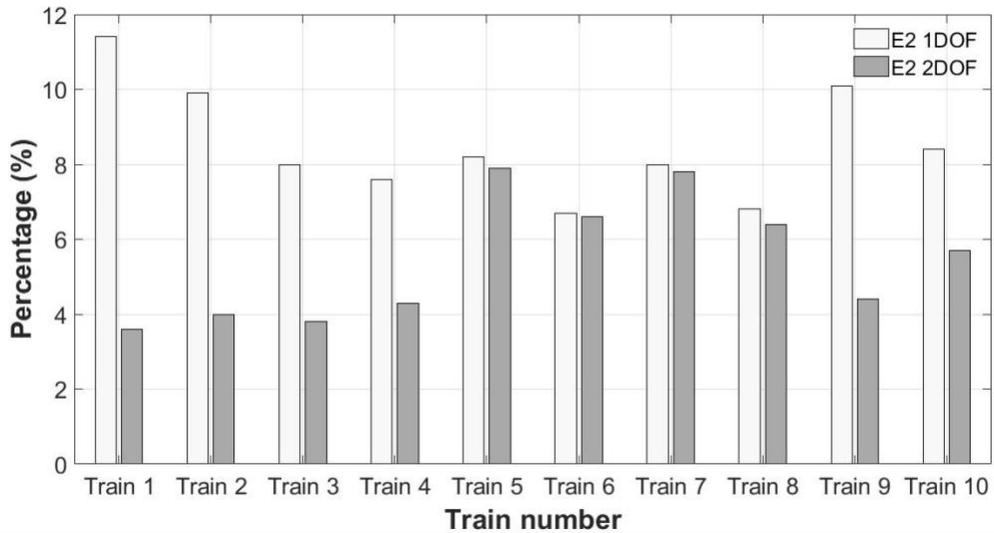

(b)

Figure 11. Comparative results between 1DOF and 2DOF; (a) Error $E_1$; (b) Error $E_2$.

## 5. Conclusion

This paper discusses the design, testing, and validation of a new condition-based monitoring method supporting management of railroad bridges. It provides railway bridges owners with more accurate total lateral displacement under trains using reference-free accelerometer sensors.

Using only two sensors per pile, this method cost-effectively increases the displacement estimation of railroad bridge piles over that obtained with existing 1DOF methods. The proposed new 2DOF method is validated through a ten laboratory experiments. In order to verify the proposed method, one column specimen was set on a shake table representing the pile under train crossing events. Ten sets of lateral bridge displacements under train-crossing events were used as inputs for the shake table to excite the specimen. The proposed 2DOF method was observed to reproduce the reference displacement with higher precision that the 1DOF method. The maximum error of $E_1\_2DOF$ corresponding to the peak error was 11.8%, and the maximum error of $E_2\_2DOF$ corresponding to the RMS error was 7.9%. The 1DOF indexes were 16.6% and 11.4%, which are 41% and 44% higher, respectively. The average value for all ten experiments of $E_2\_2DOF$ was 5.45%, which is less than 8.51% of $E_2\_1DOF$. On average, the 2DOF method reduced the error of displacement estimation by 48% and 36% in peak to peak and RMS estimations, respectively, which is a significant improvement with respect to the 1DOF method. By only adding one new sensor in each pile (two sensors total per pile), this method enables railroad bridge managers and inspectors to obtain the total transverse displacement of various piles with different foundation conditions, without knowing their structural properties. Therefore, performance based, objective management of infrastructure can be improved with higher safety and optimized investments. This model-free sensing approach will be in the future expanded to MDOF that will enable its implementation for more complex towers and buildings.

**Acknowledgments**

This study was funded by the Department of Civil, Construction and Environmental Engineering (CCEE) at the University of New Mexico; The Transportation Consortium of South-Central States (TRANSET); US Department of Transportation (USDOT), Projects No. 17STUNM02 and 18STUNM03; New Mexico Consortium Grant Award No. A19-0260-002; and "2017 Yangzhou University Graduate International Academic Exchange Special Fund Project" for providing financial support to graduate students to conduct this research. The authors declare that they have no conflict of interest.

**Data Availability Statement**

Some or all data, models, or code that support the findings of this study are available from the corresponding author upon reasonable request.

**References**

Brownjohn, J. M. (2007). Structural health monitoring of civil infrastructure. *Philosophical Transactions of the Royal Society A: Mathematical, Physical and Engineering Sciences*, *365*(1851), 589-622.

Chang, H. F. and Lin, T. K. (2019). Real-time Structural Health Monitoring System Using Internet of Things and Cloud Computing. *arXiv preprint arXiv:1901.00670*.

Dias, R. (2007). *Dynamic behaviour of high speed railway bridges. Vehicles lateral dynamic behaviour.* Dissertation for the degree of Master of Science in Civil Engineering.

Farrar, C. R., Doebling, S. W., & Nix, D. A. (2001). Vibration–based structural damage identification. *Philosophical Transactions of the Royal Society of London. Series A: Mathematical, Physical and Engineering Sciences*, *359*(1778), 131-149.

Farrar, C. R., & Jauregui, D. A. (1998). Comparative study of damage identification algorithms applied to a bridge: I. Experiment. *Smart materials and structures*, *7*(5), 704.

Farrar, C. R., & Lieven, N. A. (2007). Damage prognosis: the future of structural health monitoring. *Philosophical Transactions of the Royal Society A: Mathematical, Physical and Engineering Sciences*, *365*(1851), 623-632.

Farrar, C. R., & Worden, K. (2007). An introduction to structural health monitoring. *Philosophical Transactions of the Royal Society A: Mathematical, Physical and Engineering Sciences*, *365*(1851), 303-315.

Fisher, C. J. (2010). AN-1057 Application Note-Using an Accelerometer for Inclination Sensing. *ANALOG DEVICES, Norwood*.

Fukuda, Y., Feng, M. Q., & Shinozuka, M. (2010). Cost-effective vision-based system for monitoring dynamic response of civil engineering structures. *Structural Control and Health Monitoring*, *17*(8), 918-936.

Galvín, P., Romero, A., Moliner, E., De Roeck, G., & Martínez-Rodrigo, M. D (2021). On the dynamic characterisation of railway bridges through experimental testing. *Engineering Structures*, *226*, 111261.

Gindy, M., Vaccaro, R., Nassif, H., & Velde, J. (2008). A state‐space approach for deriving bridge displacement from acceleration. *Computer‐Aided Civil and Infrastructure Engineering*, *23*(4), 281-290.

Hannan, M. A., Hassan, K., & Jern, K. P. (2018). A review on sensors and systems in structural health monitoring: current issues and challenges. *Smart Structures and Systems*, *22*(5), 509-525.
24

Helmi, K., Taylor, T., Zarafshan, A., & Ansari, F. (2015). Reference free method for real time monitoring of bridge deflections. *Engineering Structures*, *103*, 116-124.

Hester, D., Brownjohn, J., Bocian, M., & Xu, Y. (2017). Low cost bridge load test: Calculating bridge displacement from acceleration for load assessment calculations. *Engineering Structures*, *143*, 358-374.

Huang, Q., Monserrat, O., Crosetto, M., Crippa, B., Wang, Y., Jiang, J., & Ding, Y. (2018). Displacement monitoring and health evaluation of two bridges using Sentinel-1 SAR images. *Remote Sensing*, *10*(11), 1714.

Lee, C. H., Kim, C. W., Kawatani, M., Nishimura, N., & Kamizono, T. (2005). Dynamic response analysis of monorail bridges under moving trains and riding comfort of trains. *Engineering Structures*, *27*(14), 1999-2013.

Lee, H. S., Hong, Y. H., & Park, H. W. (2010). Design of an FIR filter for the displacement reconstruction using measured acceleration in low-frequency dominant structures. *International Journal for Numerical Methods in Engineering*, *82*(4), 403-434.

Lee, J. J., Shino Lee, J. J., & Shinozuka, M. (2006). A vision-based system for remote sensing of bridge displacement. *Ndt & E International*, *39*(5), 425-431.

Moreu, F., Jo, H., Li, J., Kim, R.E., Cho, S., Kimmle, A., Scola, S., Le, H., Spencer Jr, B.F. and LaFave, J.M. (2014). Dynamic assessment of timber railroad bridges using displacements. *Journal of Bridge Engineering*, *20*(10), p.04014114.

Moreu, F., Li, J., Jo, H., Kim, R. E., Scola, S., Spencer Jr, B. F., & LaFave, J. M. (2016). Reference-free displacements for condition assessment of timber railroad bridges. *Journal of Bridge Engineering*, *21*(2), 04015052.

Moreu, F., Spencer Jr, B. F., Foutch, D. A., & Scola, S. (2017). Consequence-based management of railroad bridge networks. *Structure and Infrastructure Engineering*, *13*(2), 273-286.

Olaszek, P. (1999). Investigation of the dynamic characteristic of bridge structures using a computer vision method. *Measurement*, *25*(3), 227-236.

Ozdagli, A. I., Gomez, J. A., & Moreu, F. (2017). Real-time reference-free displacement of railroad bridges during train-crossing events. *Journal of Bridge Engineering*, *22*(10), 04017073.

Ozdagli, A. I., Liu, B., & Moreu, F. (2018). Low-cost, efficient wireless intelligent sensors (LEWIS) measuring real-time reference-free dynamic displacements. *Mechanical Systems and Signal Processing*, *107*, 343-356.

Pandey, P. C., & Barai, S. V. (1995). Multilayer perceptron in damage detection of bridge structures. *Computers & structures*, *54*(4), 597-608.

Park, J. W., Sim, S. H., & Jung, H. J. (2013). Displacement estimation using multimetric data fusion. *IEEE/ASME Transactions On Mechatronics*, *18*(6), 1675-1682.

Park, J. W., Sim, S. H., & Jung, H. J. (2014). Wireless displacement sensing system for bridges using multi-sensor fusion. *Smart Materials and Structures*, *23*(4), 045022.

Park, J. W., Lee, K. C., Sim, S. H., Jung, H. J., & Spencer Jr, B. F. (2016). Traffic safety evaluation for railway bridges using expanded multisensor data fusion. *Computer‐Aided Civil and Infrastructure Engineering*, *31*(10), 749-760.

Park, K. T., Kim, S. H., Park, H. S., & Lee, K. W. (2005). The determination of bridge displacement using measured acceleration. *Engineering Structures*, *27*(3), 371-378.

Park, S., Yun, C. B., Roh, Y., & Lee, J. J. (2006). PZT-based active damage detection techniques for steel bridge components. *Smart Materials and Structures*, *15*(4), 957.

Pregnolato, M. (2019). Bridge safety is not for granted–a novel approach to bridge management. *Engineering Structures*, *196*, 109193.

Salawu, O. S. (1997). Detection of structural damage through changes in frequency: a review. *Engineering structures*, *19*(9), 718-723.

Shin, S., Lee, S. U., Kim, Y., & Kim, N. S. (2012). Estimation of bridge displacement responses using FBG sensors and theoretical mode shapes. *Structural Engineering and Mechanics*, *42*(2), 229-245.

Spencer Jr, B. F., Moreu, F., & Kim, R. E. (2014). Structural health monitoring of railroad bridges using wireless smart sensors (WSSs): Recent real-world experiences in North America. *Life-Cycle of Structural Systems: Design, Assessment, Maintenance and Management*, 396.

Vicente, M. A., Gonzalez, D. C., Minguez, J., & Schumacher, T. (2018). A novel laser and video-based displacement transducer to monitor bridge deflections. *Sensors*, *18*(4), 970.

Wang, H., Mao, J. X., & Spencer Jr, B. F. (2019). A monitoring-based approach for evaluating dynamic responses of riding vehicle on long-span bridge under strong winds. *Engineering Structures*, *189*, 35-47.

Yoon, H., Shin, J., & Spencer Jr, B. F. (2018). Structural displacement measurement using an unmanned aerial system. *Computer‐Aided Civil and Infrastructure Engineering*, *33*(3), 183-192.

Zhu, X., & Law, S. (2007). Damage detection in simply supported concrete bridge structure under moving vehicular loads. *Journal of vibration and acoustics*, *129*(1), 58-65.